\documentclass[10.5pt,twocolumn]{article}

\usepackage[margin=0.65in]{geometry}

\usepackage{graphicx}
\usepackage{amsmath,amssymb,amsfonts}
\usepackage{textcomp}
\usepackage{array}
\usepackage{url}
\usepackage{verbatim}
\usepackage{balance}
\usepackage{caption}
\usepackage{subcaption}
\usepackage{xcolor}
\usepackage{gensymb}
\usepackage{algorithm}
\usepackage{afterpage}
\usepackage{dblfloatfix}
\usepackage{todonotes}
\usepackage{threeparttable}
\usepackage{times}
\usepackage[utf8]{inputenc}
\usepackage[colorlinks=true,linkcolor=blue,citecolor=blue,urlcolor=blue]{hyperref}

\newcommand{\arxivnotice}{%
\copyright~2024 IEEE. Personal use of this material is permitted. Permission from IEEE must be obtained for all other uses, in any current or future media, including reprinting/republishing this material for advertising or promotional purposes, creating new collective works, for resale or redistribution to servers or lists, or reuse of any copyrighted component of this work in other works.
\\ \copyright~Authors 2024. This is the author's version of the work. It is posted here for your personal use. Not for redistribution. The definitive Version of Record was published in Proceedings of the 2024 IEEE/ACM International Conference on Computer-Aided Design (ICCAD). DOI to be added upon publication.%
}

\title{\textbf{Multi-Phase Coupled CMOS Ring Oscillator based
Potts Machine} \thanks{\arxivnotice}}

\author{
    \Large Yilmaz Ege Gonul, Baris Taskin \\  
    \large Department of Electrical and Computer Engineering, Drexel University \\  
    \large Philadelphia, PA, USA \\
    \large \texttt{\{yeg26, bt62\}@drexel.edu}
}

\date{}  

\begin{document}

\maketitle






\begin{abstract}
This paper presents a coupled ring oscillator based Potts machine to solve NP-hard combinatorial optimization problems (COPs).  
Potts model is a generalization of the Ising model, capturing multivalued spins in contrast to the binary-valued spins allowed in the Ising model. Similar to recent literature on Ising machines, the proposed architecture of Potts machines implements the Potts model with interacting spins represented by coupled ring oscillators.  Unlike Ising machines which are limited to two spin values, Potts machines model COPs that require a larger number of spin values.  A major novelty of the proposed Potts machine is the utilization of the N-SHIL (Sub-Harmonic Injection Locking) mechanism, where multiple stable phases are obtained from a single (i.e. ring) oscillator.  
In evaluation, 3-coloring problems from the DIMACS SATBLIB benchmark and two randomly generated larger problems are mapped to the proposed architecture. The proposed architecture is demonstrated to solve problems of varying size with 89\% to 92\% accuracy averaged over multiple iterations. The simulation results show that there is no degradation in accuracy, no significant increase in solution time, and only a linear increase in power dissipation with increasing problem sizes up to 2000 nodes.

\end{abstract}


\maketitle

\section{Introduction}

Combinatorial optimization problems (COPs) encompass a broad spectrum of real-world applications such as logistics, drug discovery, and finance, with an increasing demand for better solvers, due to big impacts made possible with even marginal improvements in COP solutions.  With their vast search spaces, these problems require solution times increasing exponentially with increasing problem sizes using traditional computing systems \cite{COP_exponential}. A recently trending computing paradigm, Ising machines, is shown to be a strong candidate in accelerating the solution of NP-hard COPs. An Ising machine solves optimization problems by leveraging the principles of the Ising model~\cite{ising_model} from statistical physics, which describes systems of interacting spins arranged on a lattice. A COP can be encoded into an Ising machine by mapping the objective function and the constraints of the problem onto the energy function (i.e.~Hamiltonian) of the Ising model. Through the minimization of the total energy of the spin configuration of the Ising system, a solution to the mapped COP is achieved. Ising machine implementations using various technologies exist in the previous literature such as quantum Ising machines \cite{dwave}, \cite{dwave2} using qubit interactions, coherent Ising machines \cite{CIM} utilizing optics, and oscillator-based Ising machines \cite{wang_oim} using phase interactions of coupled oscillators.

The significant potential of Ising machines in solving COP problems with efficiency is a motivating factor in further research into these non Von Neumann type emerging computing architectures. In the Ising model that underlies the Ising machine implementations, such as those in \cite{dwave, dwave2, CIM, wang_oim, rtwo_ising,blim, cmos_annealer}, the spins of interacting nodes are modeled with 2 spin values.  A generalized model of the Ising model is the Potts model~\cite{potts_model}, where the same physics principles of energy minimization model the interaction between multi-valued spins. In terms of mapping COPs, the Potts model addresses the limitation of Ising machines stemming from the limitation to 2 allowed spin values. For some COPs, such as a max-cut problem, the mapping to an Ising machine is seamless, as 2 spins are sufficient to model the relationship between the optimization goals of the max-cut problem and the Ising Hamiltonian. On the other hand, the mapping of other COP problems, such as a max-K cut or a graph coloring problem, demands sophisticated mapping approaches \cite{many_ising} or leads to inefficient usage of resources.  

This paper proposes a CMOS ring oscillator (ROSC) based Potts machine (ROPM) capable of mapping and solving COPs requiring multivalued spins. The proposed ROPM obtains multivalued spins from a single oscillator, enabling the mapping of such COPs using the Potts formulation as detailed in Section 3.1, instead of the Ising formulation that only directly maps 2-spin problems without the insertion of auxiliary spins and overhead in the graph size \cite{many_ising}. The proposed ROPM architecture leverages the existing coupled-ROSC based Ising machine architecture in~\cite{prob_fabric}.  A major enabler of the proposed ROSC-based Potts architecture, and a significant differentiator from the ROSC-based Ising machine in literature, is the $N^{th}$ order Sub-Harmonic Injection Locking (N-SHIL) \cite{seal_potts} mechanism. The N-SHIL provides the stimulus to the coupled ROSCs necessary to assume multiple distinct phases of the ROPM at each individual oscillator. 
A wide range of simulation-based studies are performed on the selected case study of the 3-coloring problem, achieving solutions with 89\% - 92\% accuracies averaged over multiple iterations with varying problem sizes. 

The rest of this paper is organized as follows. In Section~2, the literature on ROSC-based Ising machines is summarized. This section includes a general technical background on Ising models, as well as the details of the ROSC based Ising machine in the literature that is leveraged in this work.  In Section~3, the proposed coupled ROSC based Potts machine~(ROPM) is introduced.  
In Section~\ref{sec:simulations}, simulation results are presented for DIMACS SATLIB benchmarks of 3-coloring graphs, as well as larger COP problems to show the efficiency and scalability of the proposed ROPM. Concluding remarks are presented in Section~\ref{sec:conclusions}.

\section{Background on ROSC based Ising Machines}
\label{sec:ising}
The technical background of general Ising model computation is summarized in Section~\ref{sec:theory_ising}.  The literature review of existing ROSC-based  Ising machines (ROIM), as they relate to the proposed ROPM, is presented in  Section~\ref{sec:litreview_ising}. 

\subsection{General Theory on Ising Model and Machines}
\label{sec:theory_ising}

Ising model Hamiltonian $H(s)$ \cite{ising_model} captures the energy of the interacting spins in a generalized Ising system:
\begin{equation}
  H(s) = \sum_{i,j} J_{ij} \cdot s_is_j,
\end{equation}
where $s_i \in \{-1,+1\}$ is a binary valued spin on the $i^{th}$ node on the model and $J_{ij}$ is the coupling strength between connected spins $s_i$ and $s_j$. 

Coupled CMOS ring oscillators, same as other types of self-sustained oscillators, represent interacting Ising spins thanks to a physical phenomenon called injection locking \cite{injection_lock}. Through injection locking, oscillators are able to phase lock to an external signal with a frequency close to the natural oscillation frequency of the oscillators when coupled with a conducting medium.   The injection locking of neighboring oscillators is akin to the interaction of spins in the Ising model, where the neighboring spins resolve to a low energy state defined by the Ising Hamiltonian.  
In an oscillator-based Ising machine~(OIM)~\cite{wang_oim},
two distinct phase values of $\theta_{s_i}$ and  $\theta_{s_j}$, ideally $180 \degree$ apart (0 \degree, 180\degree), represent the binary spins in the Ising Hamiltonian $H(\theta_{s})$ :
\begin{equation}
  H(\theta_{s}) = \sum_{i,j} J_{ij} \cdot cos(\theta_{s_i} - \theta_{s_j}),
\end{equation}
The strength of the coupling medium $J_{ij}$ is arbitrated through a mechanism such as device or interconnect sizing between the oscillators.

\subsection{Review of ROSC based Ising Machines }
\label{sec:litreview_ising}

\begin{figure}[t]
    \centering
    \includegraphics[width=\linewidth]{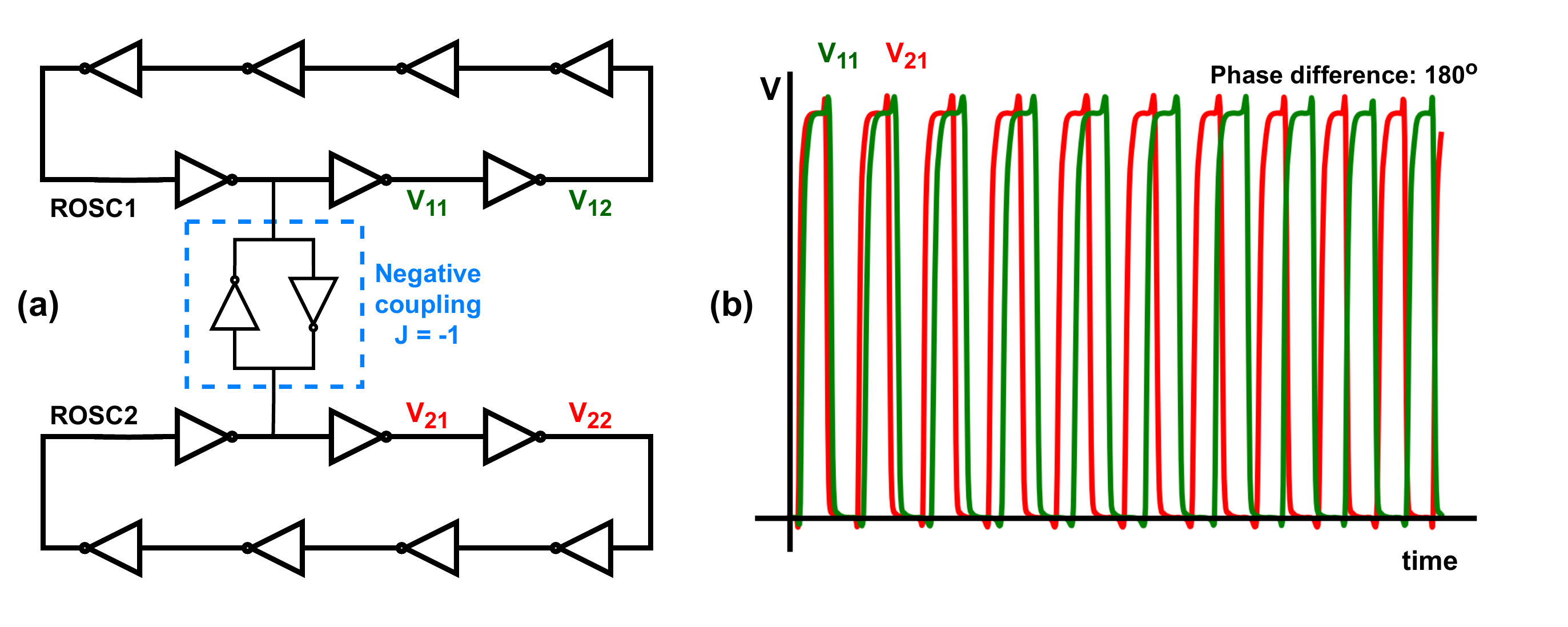}
     \caption{a) Two ROSCs negatively coupled with an inverting coupling medium (B2B inverters) b) Coupled ROSC phases progressively locking out of phase through time}
    \label{fig:coupled_roscs}
\end{figure}

In the recent literature \cite{chris_kim_560, prob_fabric, chris_kim_nat}, coupled-ROSC based Ising machines~(ROIMs) are demonstrated on silicon, solving max-cut type COP problems. 
Figure \ref{fig:coupled_roscs}(a) shows two negatively coupled ROSCs, representative of larger ROIMs that could have more number of ROSCs (i.e. spins). 
When two ROSCs are negatively coupled, as in Figure~\ref{fig:coupled_roscs}(a), their phases settle out of phase as demonstrated with the waveforms in Figure \ref{fig:coupled_roscs}(b). A coupling medium that is non-inverting (positive coupling $J_{ij}=1$) or inverting (negative coupling $J_{ij}=-1$) is used to couple the ROSCs, or a gated coupling medium (disabled B2B inverter) is used to represent no coupling ($J_{ij}=0$) between nodes.  

In the mapping of a max-cut problem, as theoretically proposed in~\cite{wang_oim} and implemented in silicon in~\cite{prob_fabric}, each node maps to a ROSC, and each graph edge maps to gated B2B inverters for coupling. The mapping is straightforward, yet, when more than two oscillators are coupled together, phase contention occurs due to contradicting forces acting on a single oscillator phase. Under phase contention, oscillator phases can settle in intermediary values, instead of the discretely valued (Ising) phases.
The strength of the couplings is adjusted ~\cite{prob_fabric} so that the couplings are strong enough to couple the ROSCs but weak enough to allow the ROSCs to keep oscillating in the presence of phase contentions.

\begin{figure*}[t]
     \centering
     \begin{subfigure}[]{0.11\linewidth}
         \centering
         \includegraphics[width=\linewidth]{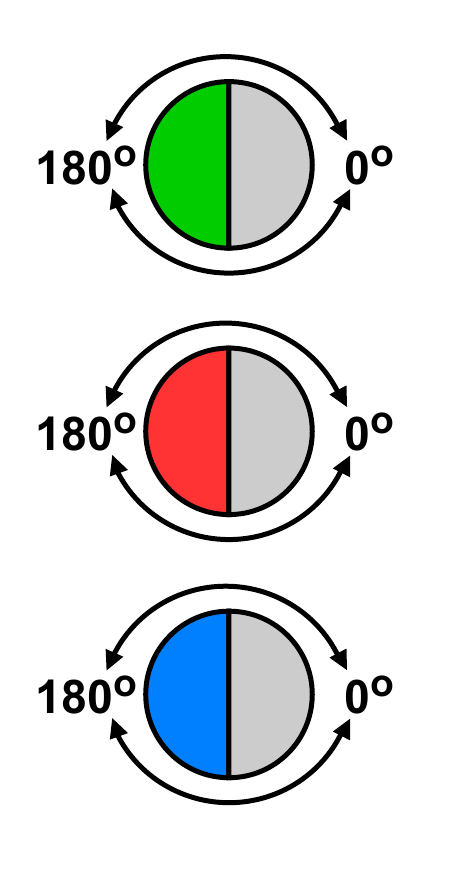}
     \end{subfigure}
     \begin{subfigure}[]{0.342\linewidth}
         \includegraphics[width=\linewidth]{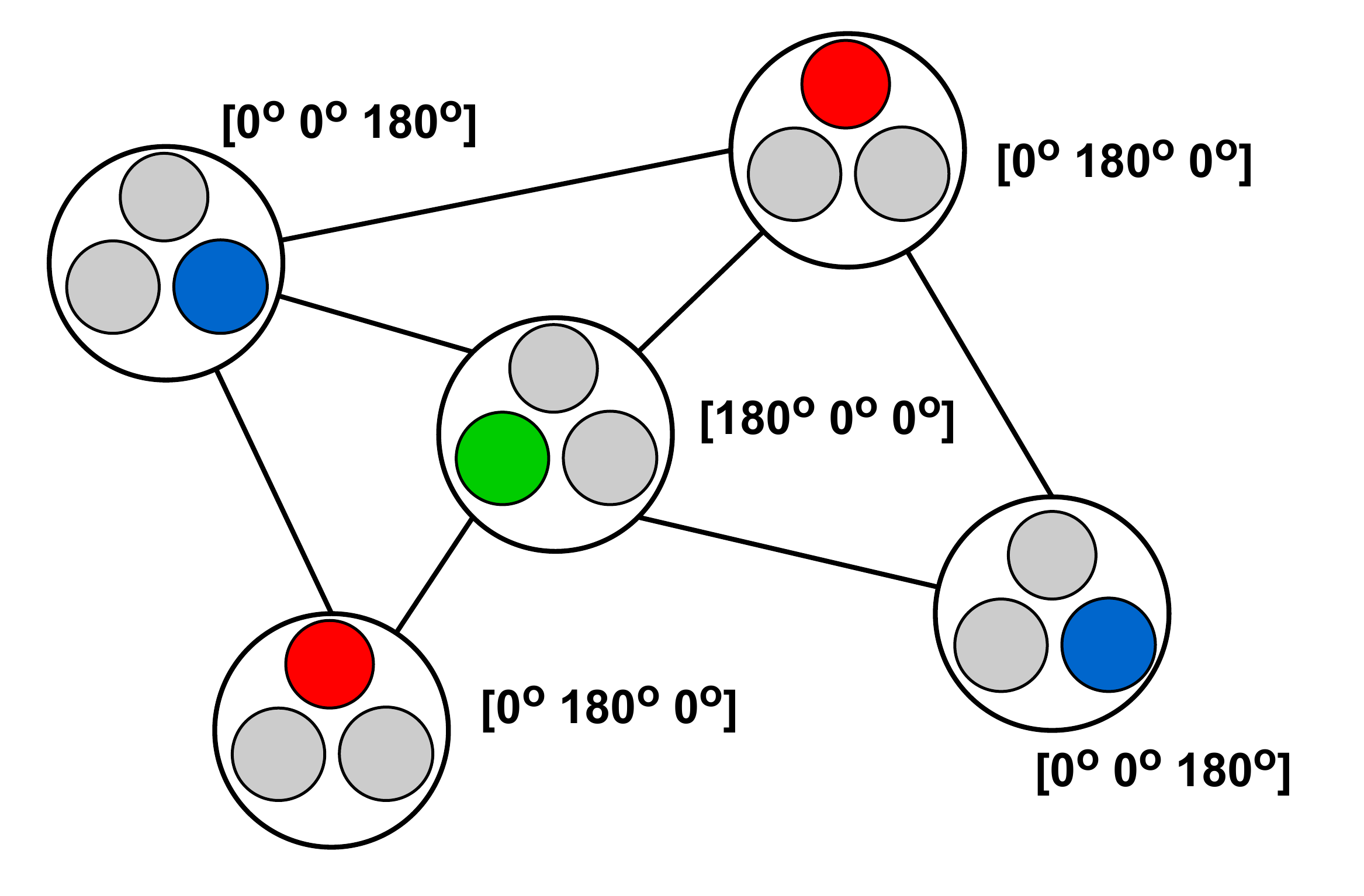}
         \caption{}
     \end{subfigure}
     \begin{subfigure}[]{0.11\linewidth}
         \includegraphics[width=\linewidth]{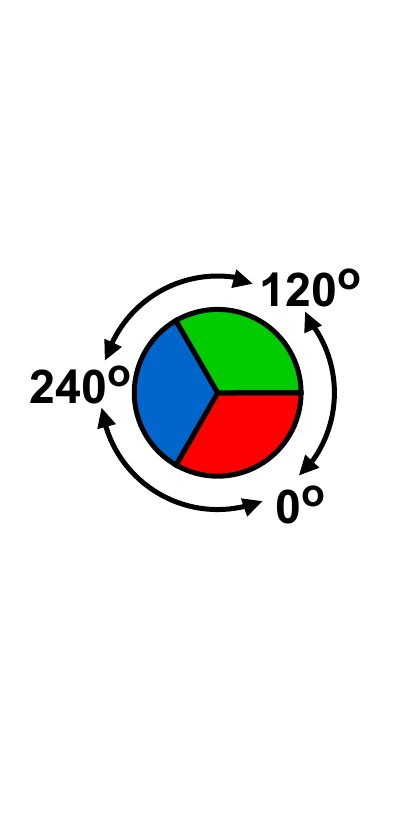}
     \end{subfigure}
     \begin{subfigure}[]{0.342\linewidth}
         \includegraphics[width=\linewidth]{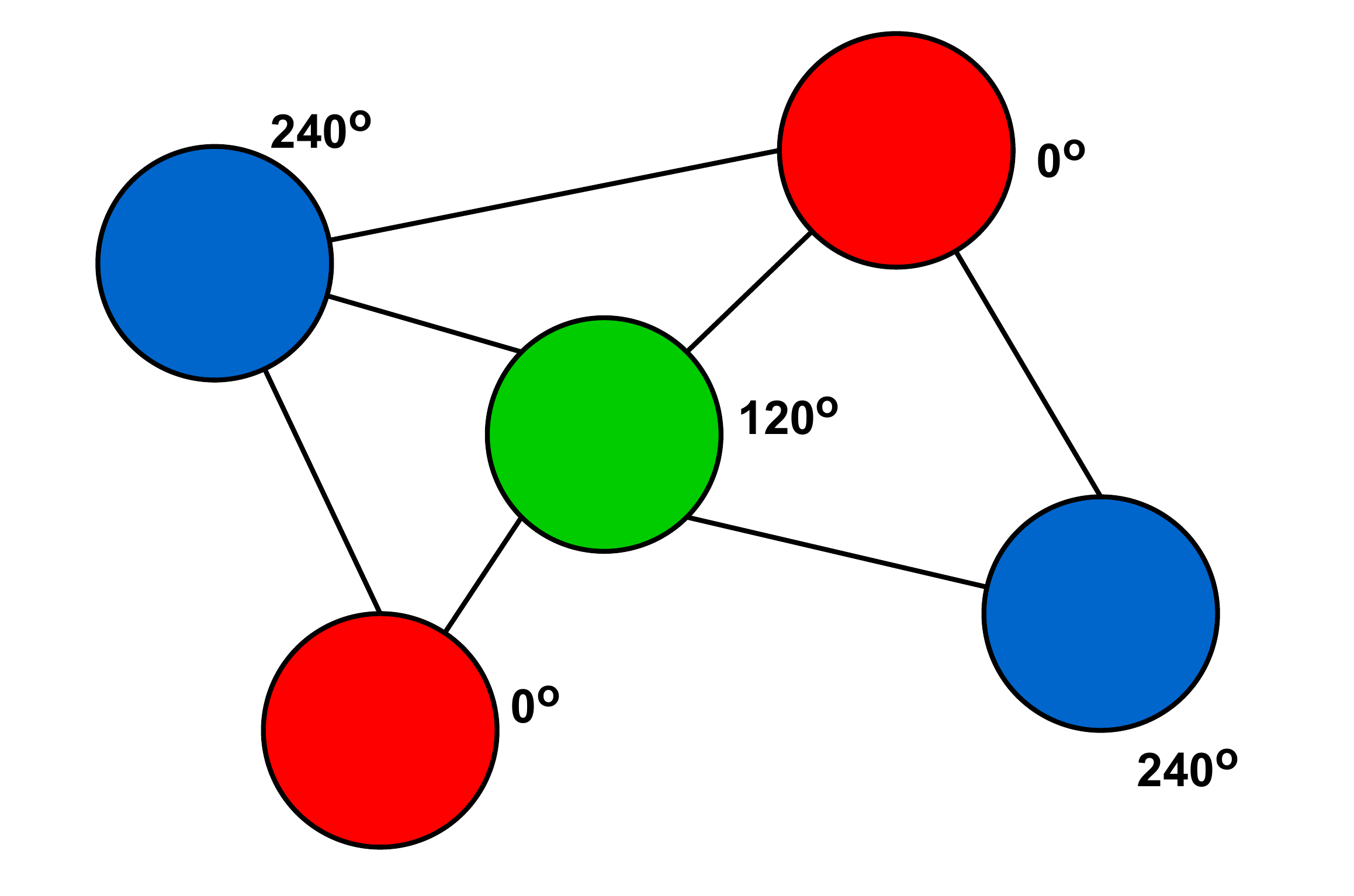}
         \caption{}
     \end{subfigure}
     \caption{A 5-node graph 3-colored with a) an OIM limited to capturing 2 distinct spins with a single oscillator b) an OPM capable of capturing 3 distinct spins with a single oscillator}
    \label{fig:2vs3}
\end{figure*}

A method in literature to resolve phase contentions in ROIMs is sub-harmonic injection locking~(SHIL). SHIL is utilized in~\cite{wang_oim} and~\cite{chris_kim_nat}, where the oscillator nodes of the Ising machines are binarized, i.e.  (+1,-1), specifically $180 \degree$ apart (0 \degree, 180\degree). It is mathematically and empirically shown in~\cite{wang_oim} that the injection of SHIL to the oscillators in the system significantly increases the accuracy of Ising machine solutions.

\section{Coupled ROSC-based Potts Machine Design}
\label{sec:potts}
\label{sec:ising}
Technical background of general Potts model computation is summarized in Section~\ref{sec:theory_potts}.  The literature review of existing implementations of Potts machines, as they relate to the proposed ROPM, is presented in  Section~\ref{sec:litreview_potts}.  Design principles and operation of the proposed ROPM are categorically presented in Sections~\ref{sec:rosc_potts} through~\ref{sec:probabilistic_potts}.

\subsection{General Theory on Potts Model and Machines}
\label{sec:theory_potts}

Potts model~\cite{potts_model} is the generalization of the Ising model, describing the interactions of spins that can take multiple values.  
Consequently, Potts machines are prime to solve COPs that cannot be modeled natively with 2 spins (i.e. with an Ising machine) more effectively.  
The standard Potts model Hamiltonian $H_{Potts}$ is: 
\begin{equation}
\large
  H_{Potts} = \sum_{i,j} J_{ij} \cdot \delta (s_i,s_j),
\end{equation}
where $s_i \in \{0,1,2, ...., N-1\}$  is an N-valued spin on the
$\textit{i}^{th}$ node of the model, and $J_{ij}$ is the interaction strength between the spins $s_i$ and $s_j$.

In the case of an Oscillator-based Potts machine~(OPM), the Hamiltonian (also called vector Potts Hamiltonian) becomes:

\begin{equation}
\large
  H_{v_{Potts}}(\theta_{s}) = \sum_{i,j} J_{ij} \cdot cos(\theta_{s_i} - \theta_{s_j}),
\end{equation}
where $H_{v_{Potts}}$ is the vector Potts Hamiltonian and the $\theta_{s_i} - \theta_{s_j}$ term indicates the phase difference between the $\textit{i}^{th}$ and $\textit{j}^{th}$ oscillator.  For an N-phase Potts machine, the oscillator phases $\theta_{s_i}$ are limited (quantized) to N phases equally spaced within $[0,2\pi]$ range: 
\begin{equation}
\large
  \theta_{s_i} =  \frac{2\pi s_i}{ N},   s_i \in \{ 0, 1, 2, ... N - 1\}
  \label{eq:pottsphase}
\end{equation}

When the Potts machine is executed, the spins of nodes gradually resume values that minimize the Hamiltonian much like the Ising machines. 
In the ideal scenario, the ground (i.e.~minimal energy) state of the Potts machine is reached, which corresponds to the optimal solution of the COP.

In~\cite{many_ising}, COPs requiring multivalued spins have existing 2-spin Ising formulations where multiple binary spins are used to represent a single multivalued spin. The Ising model becomes inefficient in mapping and solving COPs as the required spin values increase in number. Potts model can capture these problems with a smaller Hilbert space than that of the Ising models~\cite{many_ising}.  
The main defining advantage of OPMs over OIMs, in being able to represent more than 2 spins by a single oscillator, is exemplified in Figure~\ref{fig:2vs3}. Figure \ref{fig:2vs3}(a) shows the mapping of a 5-node 3-coloring problem with an Ising machine representing each color with a separate binary-valued spin for every vertex. For the 3-coloring problem considered in Figure \ref{fig:2vs3}(a), only 1 out of 3 spins can be +1 ($180 \degree$) at a time representing the color of that vertex in a one-hot scheme. Figure \ref{fig:2vs3}(b) illustrates the use of a single oscillator to represent the whole set of 3 colors on a vertex. 
For the 5-node 3-coloring problem, the OPMs lead to 66\% less number of oscillators (5 oscillators in OPM vs 15 oscillators in OIM), as well as reductions in couplings and associated control. The improvements are more pronounced for higher K values of K~-~coloring problems.  
An OPM built by individual oscillators capable of representing K distinct spins would drastically improve the energy efficiency, speed of the system, and ease of mapping to that OPM system.

\subsection{Review of Potts Machines in Literature vs the Proposed ROPM}
\label{sec:litreview_potts}
Potts machine implementations using non-CMOS technologies are proposed in previous literature  \cite{optical_potts}, \cite{coherent_potts} optical, and, \cite{potts_nature} hybrid (optical \& digital). Also called coherent Potts machines (CPM), these implementations use optical laser pulses as spins, requiring complex fabrication processes and up to a kilometer-long optical fibers. 
In contrast, proposed ROPM can be fabricated with a lower cost in a much more miniaturized size leveraging the advantages of silicon. An Ising and Potts annealer, using single photon avalanche-diodes that is compatible with CMOS technology, is also previously implemented \cite{cmos_compatible_potts}. Comparison of the proposed ROPM and other Potts machines in terms of other performance metrics are discussed in more detail in Section~\ref{sec:comparison}.

\subsection{ROSCs for the Proposed ROPM}
\label{sec:rosc_potts}

The proposed ring oscillator based Potts machine (ROPM) is implemented using a coupled ROSC network similar to the one proposed in \cite{prob_fabric}. 
CMOS ring oscillators, compared to the other types of self-sustained oscillators, are more energy efficient, more area efficient and lower cost to manufacture \cite{prob_fabric}. Also compared to other components that are used in Ising machine implementations to represent spins such as latches \cite{blim} or SRAM cells \cite{cmos_annealer} that can stabilize at two distinct points, oscillators can represent a continuous range of phases and be stabilized (through N-SHIL) at any number of discrete points that are practically possible. On the other hand, ROSCs are more susceptible to process variations and have higher jitter (phase noise).  Complicated systems, or systems that necessitate stability and accuracy, are not well tailored with implementation with ROSCs. In contrast, Ising machines are intrinsically tolerant to and can potentially benefit from noise. Similar arguments hold true for the proposed ROPM. 
Another advantage of the simple design factor is in being amenable to sizing, without high overhead, to accommodate the N-SHIL operation. 

\begin{figure}[]
   \centering
   \includegraphics[width=\linewidth]{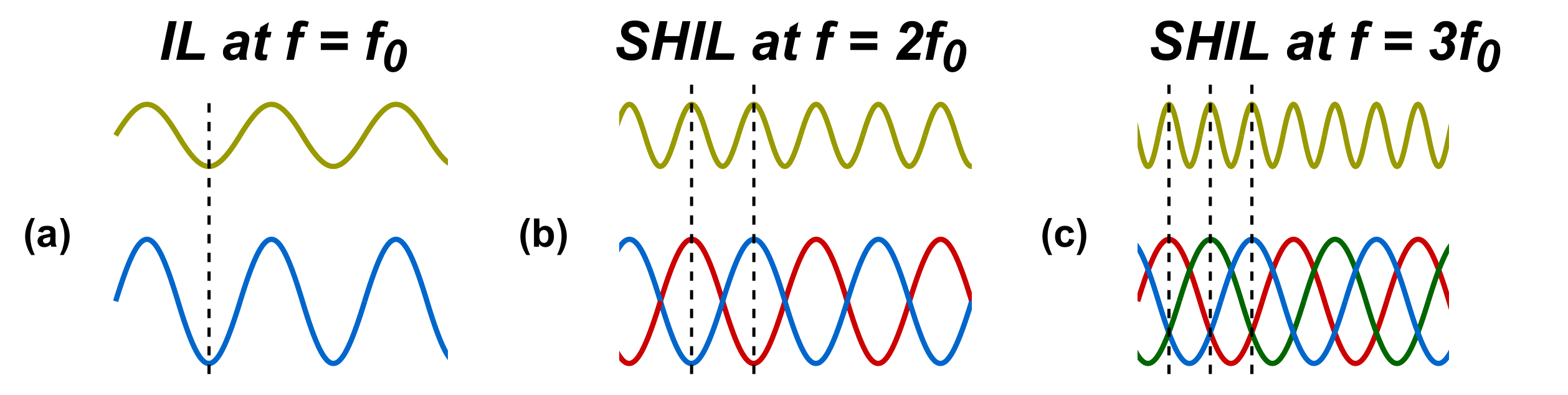}
    \caption{Illustrations of phase locking in the cases of perturbations by SHIL with a)~the $1^{st}$ harmonic (fundamental) b)~$2^{nd}$ harmonic c)~$3^{rd}$ harmonic of the base frequency }
   \label{fig:n-shil}
\end{figure}

\subsection{N-SHIL for the Proposed ROPM}
\label{sec:shil_potts}
Other than the fundamental harmonic of an oscillator, oscillators can phase lock into signals with a frequency that is an exact integer multiple (harmonic) of the natural frequency of the oscillator frequency, through SHIL~\cite{analysis_shil} as illustrated in Figure~\ref{fig:n-shil}.  
An oscillator perturbed by SHIL having $N^{th}$ multiple of its natural frequency, becomes phase-locked into one of $N$ possible stable phases that are equally spaced in the range [0,2$\pi$]. 
This is used as the enabling mechanism for the N-valued spins in the proposed ROPM architecture.
The existence of N-phase optimum solutions in a system of coupled oscillators in the presence of N-SHIL is mathematically proven in \cite{seal_potts}.

The proposed ring oscillator based Potts machine (ROPM) is implemented utilizing 3-SHIL (3 discrete phases for 3-coloring), in contrast to the 2-SHIL utilized for (i.e. 2 phase) Ising machines in~\cite{prob_fabric}.  Multi-valued spin representations $\theta_{s_i}$ of the proposed ROPM are achieved with individual ROSCs, each receiving a 3-SHIL synchronization stimuli (denoted as SYNC in) Figure~\ref{fig:blocks}(a).  
In effect, each ROSC locks in phase to one of the 3 possible stable phases that are 120 \degree spaced in the range $[0,2\pi]$, as depicted by Equation \ref{eq:pottsphase}.

\subsection{Implementation of the Proposed ROPM }
\label{sec:pottsimplementation}

\begin{figure}[b]
     \centering
     \begin{subfigure}[]{0.6\columnwidth}
         \centering
         \includegraphics[width=\linewidth]{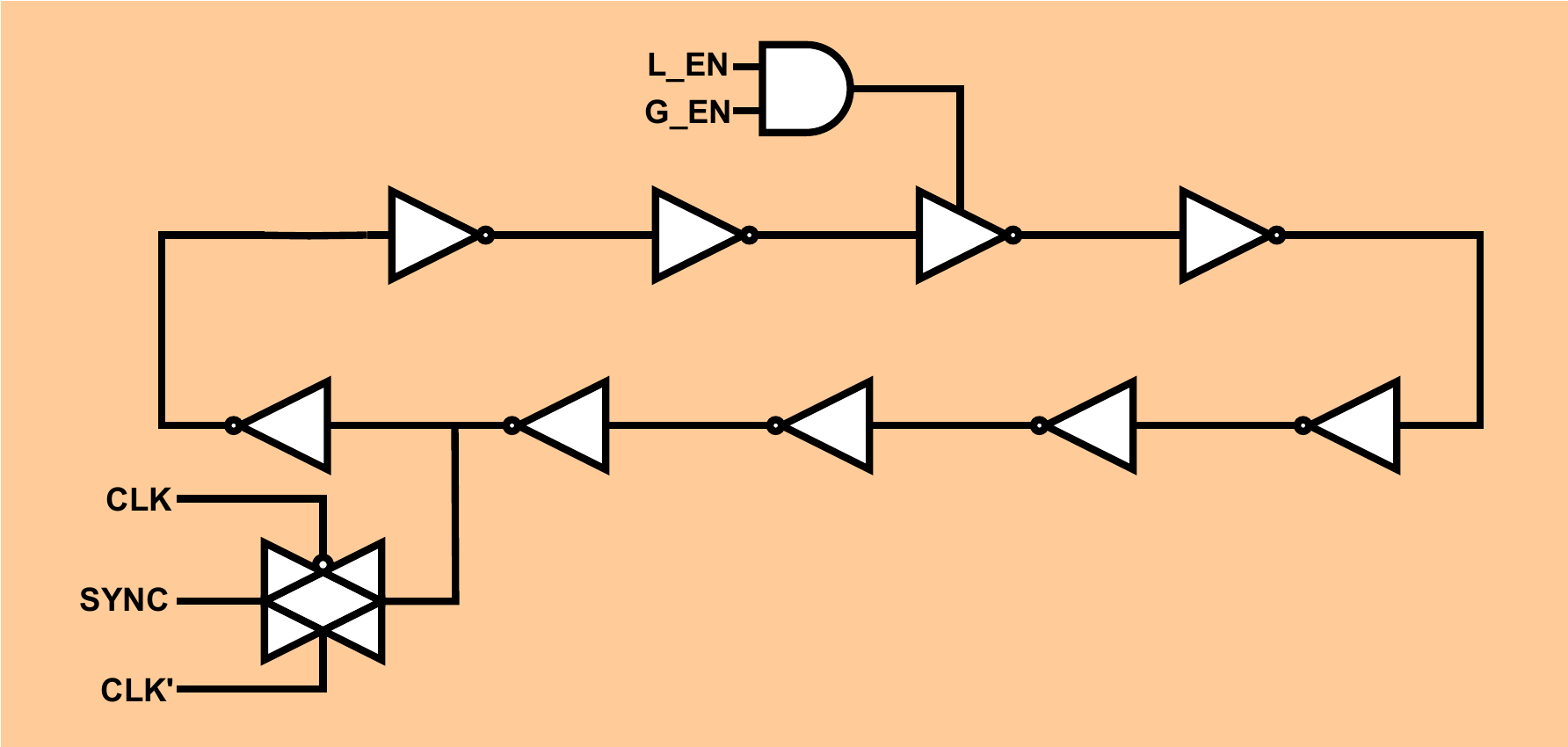}
            \caption{}
     \end{subfigure}
     \hfill
     \begin{subfigure}[]{0.35\columnwidth}
         \includegraphics[width=\linewidth]{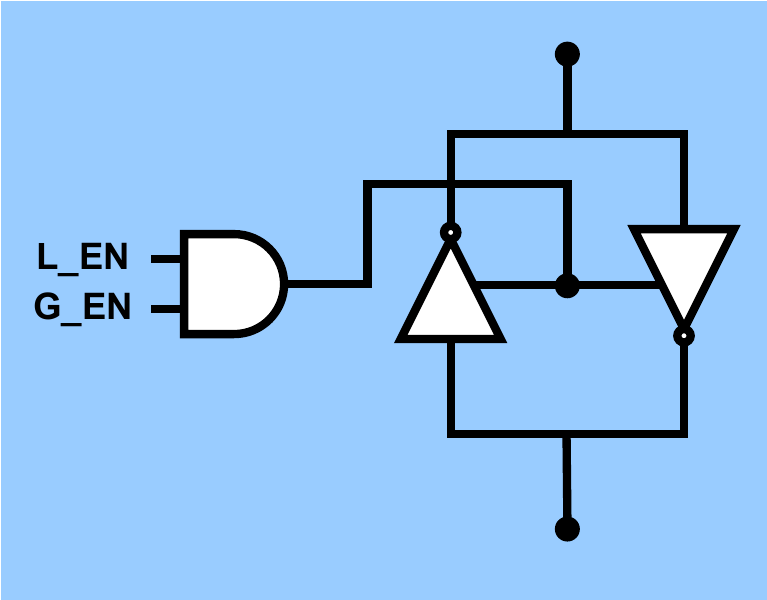}
         \caption{}
     \end{subfigure}
     \caption{a) ROSC block with local and global enable signals (L\_EN and G\_EN) and SYNC (for N-SHIL) injected through a pass transistor b) Coupling block containing B2B inverters with local and global enables (L\_EN and G\_EN)
     }
    \label{fig:blocks}
\end{figure}

The ROPM is built using the ROSC and coupling blocks as shown in Figures~\ref{fig:blocks}(a) and~\ref{fig:blocks}(b) respectively. A 9-stage ROSC implementing the proposed ROPM is designed 
oscillating with a frequency at around 7 GHz. 
Although slower ROSCs at the cost of device complexity can be preferred for more robustness against PVT variations~\cite{prob_fabric}, this frequency is selected to investigate the accuracy of ROPM operation at a challenging design settings at the GHz range. 
Similar to \cite{prob_fabric}, back-to-back~(B2B) inverters as shown in Figure~\ref{fig:blocks}(b) are used as the coupling components. The inverting feature of B2B is suitable for the graph coloring problem, where neighboring ROSCs are required to assume different phases to satisfy the coloring objective. Phase interactions in the coupled ROSC network move the system to lower energy states under tuned operating conditions where the coupling strength of the B2Bs are neither too strong to stop the oscillation of the coupled ROSCs nor too weak to perform injection locking \cite{prob_fabric}. The SYNC perturbation signal or 3-SHIL [shown in Figure~\ref{fig:blocks}(a)] works in tandem with the couplings to converge to 3-phase ground states. The presence of the 3-SHIL connected to ROSCs constrains the ROPM to only converge to 3-phase ground states.

The SYNC signal in Figure~\ref{fig:blocks}(a) is generated as a sine wave with $N^{th}$ multiple of the base frequency with a lower amplitude than the base oscillations of ROSCs.
The SYNC signal carrying the N-SHIL stimuli interacts with the ROSCs in the ROPM through a coupling medium similar to the couplings between ROSCs. In this work, SYNC is injected to each ROSC through a pass-transistor as illustrated in Figure~\ref{fig:blocks}(a). Alternative coupling components can be also used such as a single NMOS transistor in prior work~\cite{chris_kim_nat} to connect a SYNC signal carrying 2-SHIL to the ROIM. The clock input denoted as CLK can be used for turning SYNC on and off, to create annealing schedules to increase accuracy \cite{chris_kim_nat}. For SYNC to seamlessly lock the oscillator phases into N equally spaced discrete phases, the relative strengths of the SHIL injection and the couplings (by sizing the pass-transistors accordingly), and the amplitude of the SYNC are critical to tune. When the SYNC signal is relatively too weak, ROSC phases do not get discretized into the desired 3 phases. Alternatively, when the SYNC signal is relatively too strong, the 3-SHIL undesirably dominates the self-oscillations of the local ROSCs.

\begin{figure}[]
     \centering
     \begin{subfigure}[]{0.45\columnwidth}
         \centering
         \includegraphics[width=\linewidth]{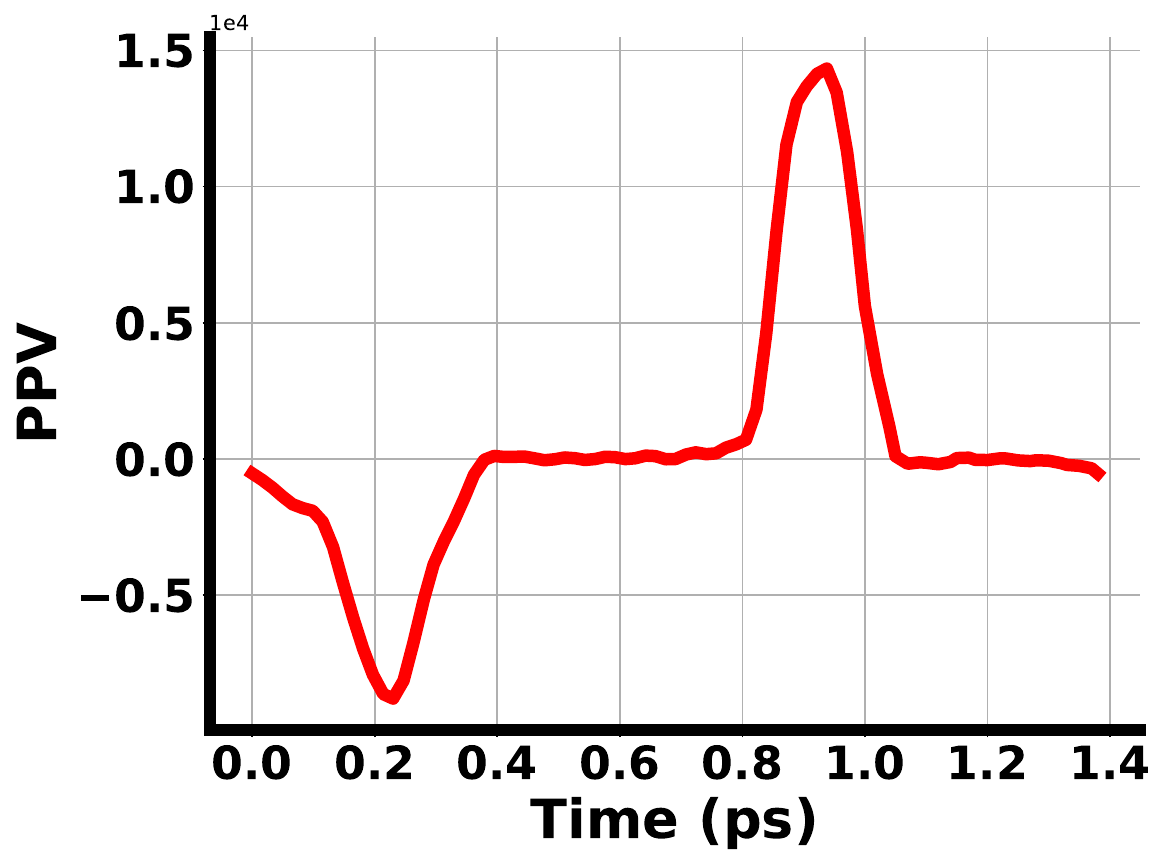}
            \caption{}
     \end{subfigure}
     \hfill
     \begin{subfigure}[]{0.45\columnwidth}
         \includegraphics[width=\linewidth]{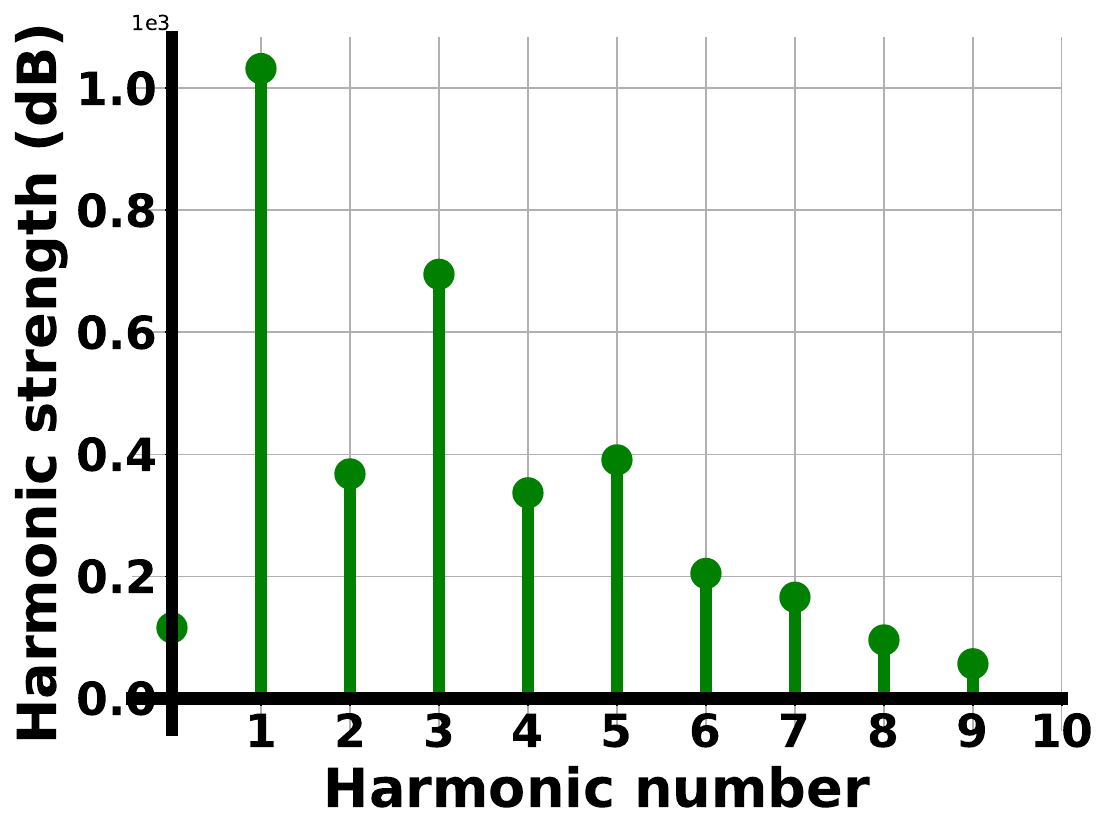}
         \caption{}
     \end{subfigure}
     \caption{3-SHIL susceptibility analysis of the proposed ROSC sizing  (a) PPV waveform from one the ROSC nodes b) Fourier transform of the PPV showing harmonic components}
    \label{fig:ppv}
\end{figure}

\begin{figure*}[t]
     \centering
     \begin{subfigure}[]{0.33\linewidth}
         \centering
         \includegraphics[width=\linewidth]{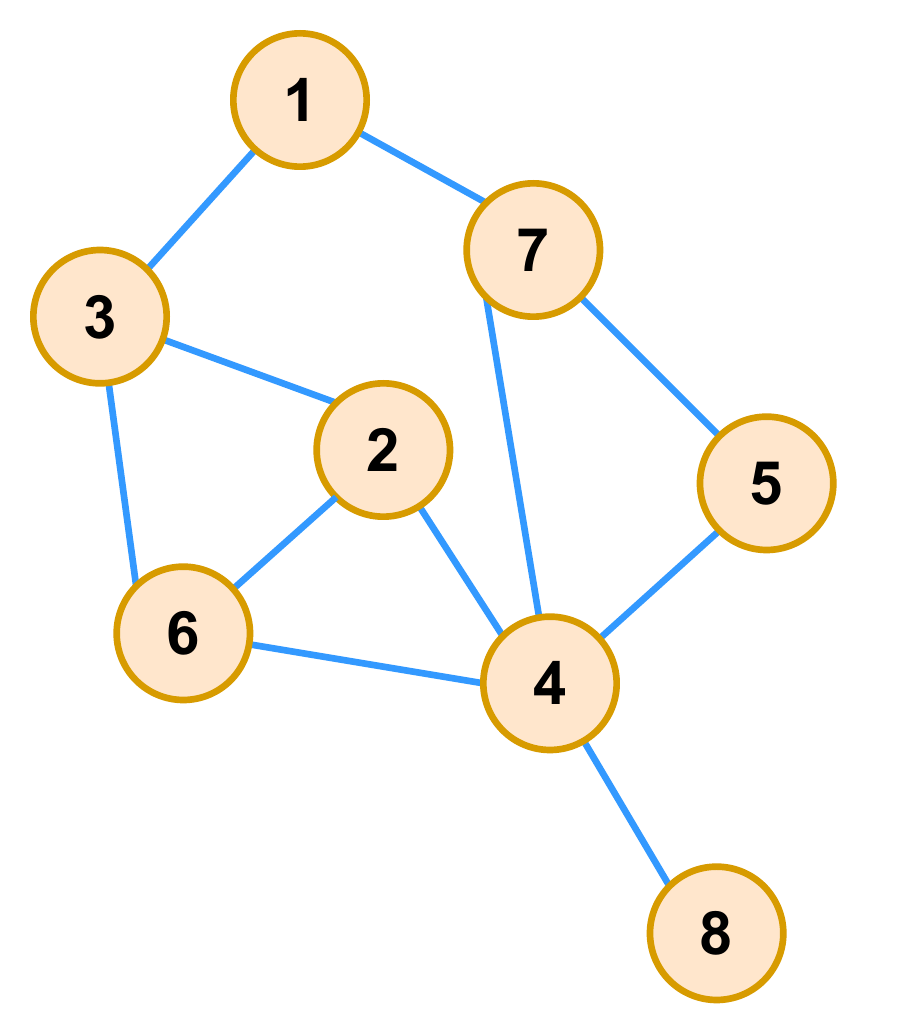}
            \caption{}
     \end{subfigure}
     \hfill
     \begin{subfigure}[]{0.62\linewidth}
         \includegraphics[width=\linewidth]{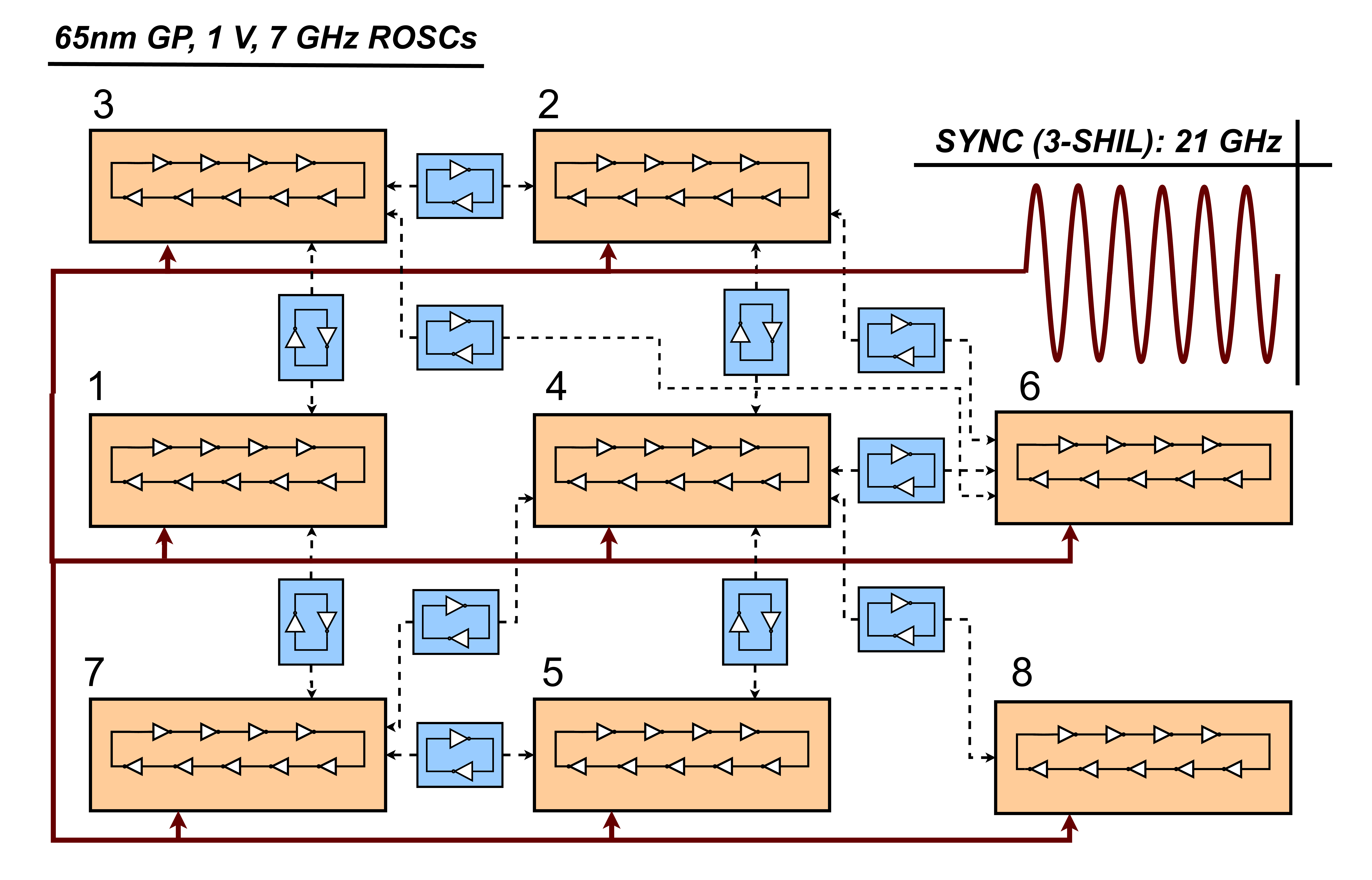}
         \caption{}
     \end{subfigure}
       \begin{subfigure}[]{0.4\linewidth}
         \includegraphics[width=\linewidth]{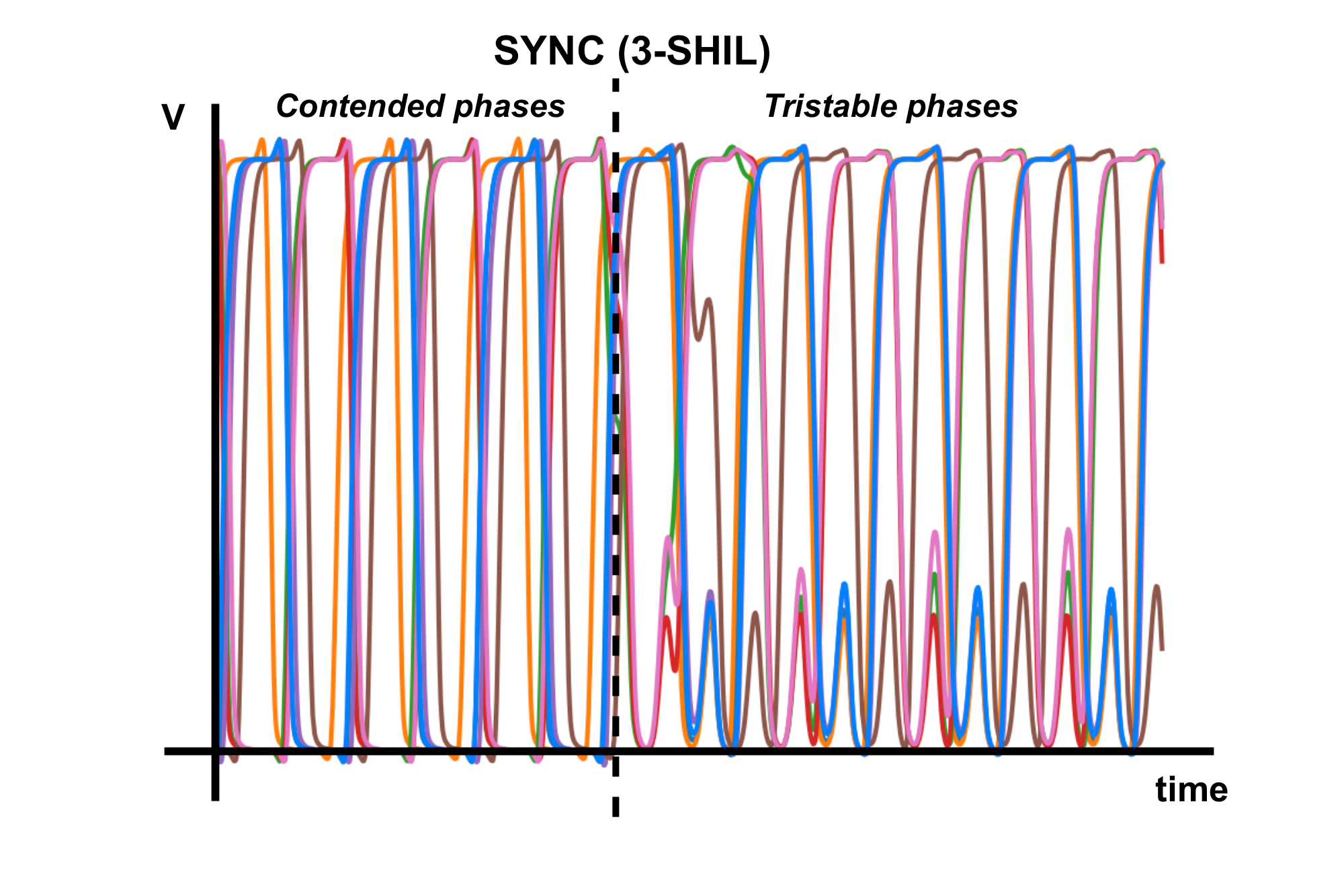}
         \caption{}
     \end{subfigure}
          \begin{subfigure}[]{0.53\linewidth}
         \includegraphics[width=\linewidth]{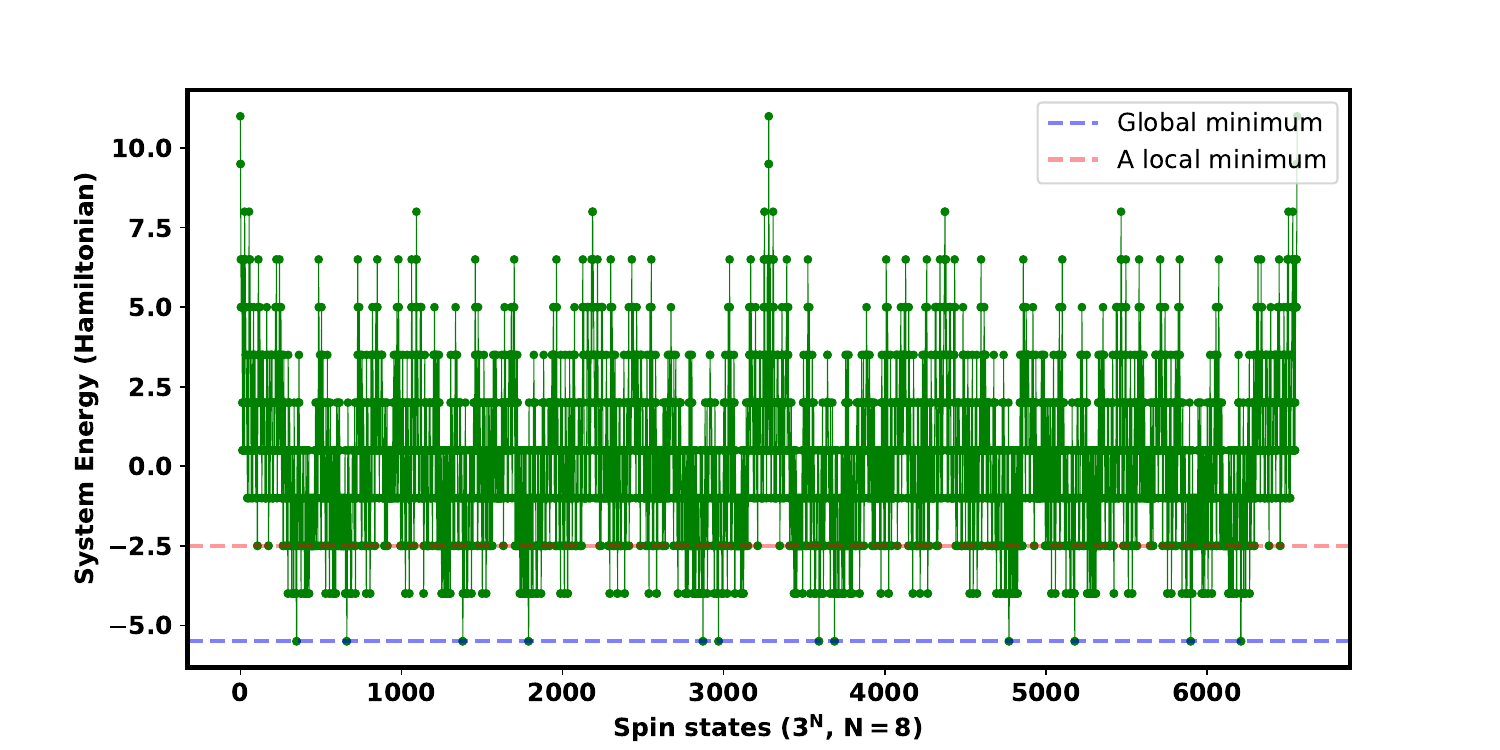}
         \caption{}
     \end{subfigure}
     \caption{a) A 8-node 3 colorable graph b) Graph in (a) mapped to an 3-phase ROPM c) Waveforms demonstrating the ring oscillator phases locking into tristability with SYNC d) Search space of the 8-node problem with $3^8$ possible phase combinations} 
      \label{fig:opm}
\end{figure*}

Both ROSCs and couplings are controlled by one global enable signal G\_EN and a local enable signal L\_EN, as shown in Figures~\ref{fig:blocks}(a) and~\ref{fig:blocks}(b), providing programmability. Local enable signals L\_EN are used to selectively turn on ROSCs and B2B couplings to map a particular problem to the ROPM. Global enable signal G\_EN is used to turn on all of the ROSCs and couplings simultaneously.  In terms of the sequence of operations in the proposed ROPM, the neighboring ROSCs are first coupled without active N-SHIL, initializing phase interactions between the ROSCs of the ROPM.  Shortly after, 3-SHIL is activated (SYNC in Figure~\ref{fig:blocks}) to discretize the indecisive phases resulting from phase contentions in the ROPM into 3 distinct phases.

The sizing of the ROSC in Figure~\ref{fig:blocks}(a) is also one of the enablers of the ROPM, making the interaction with the 3-SHIL possible. For an $N^{th}$ harmonic SHIL to seamlessly interact with an oscillator, the oscillator must be designed as susceptible to such a perturbation~\cite{analysis_shil}, \cite{wang_metronome}. Susceptibility of an oscillator to an $N^{th}$ harmonic SHIL can be observed through the Perturbation Projection Vector (PPV) waveform~\cite{analysis_shil} obtained by the Spectre $pss$ analysis, shown in Figure~\ref{fig:ppv}. The strength of $N^{th}$ harmonic component can be observed  at the Fourier transform of the PPV. In order to strengthen different harmonic components of the PPV, the oscillator design is modified accordingly. In the case of ROSCs, relative sizing of the NMOS and PMOS transistors i.e. the symmetry of the inverters, boosts different harmonic components. For example in order to boost the $2^{nd}$ harmonic of the PPV, inverters must be designed asymmetrically, whereas in order to boost the $3^{rd}$ harmonic, symmetrical inverters are designed. In the ROSCs implementing the ROPM in this work, NMOS and PMOS widths are set up with $1:1$ ratio. Fourier transform of the PPV shown in Figure \ref{fig:ppv}(b) illustrates that the $3^{rd}$ harmonic component is stronger than the other harmonics, thus the ROSC becomes more susceptible to 3-SHIL.

\subsection{Mapping 3-Coloring Problem to a Potts Machine}
\label{sec:map_potts}

Graph coloring is one of the most extensively studied COPs due to its wide-ranging applications in diverse domains including network design, and scheduling algorithms.
Given a simple graph
G(V, E), graph coloring, also called K-coloring, consists of assigning
one of K colors to every vertex in such a manner that no
adjacent vertices are assigned the same color. Graph coloring problem is selected to demonstrate the proposed ROPM, due to its direct mapping to a multi-valued spin-interaction system~\cite{higher_order}. For simplicity of demonstration, the 3-coloring problem is mapped to ROPM that utilizes the 3-SHILs detailed in Figure~\ref{fig:n-shil}, and the ROPM modeling Equation~\ref{eq:pottsphase} setting N = 3.  The 3-coloring problem is a constrained K-coloring problem where K=3, which is still NP-complete. In the more general case, the proposed Potts solver methodology can be extended to other well-known COPs with some effort as well, since most COPs are reducible to each other~\cite{reducibility}.

Figure \ref{fig:opm} exemplifies the operation of the ROPM mapping a random 8-node and 11-edge 3-colorable graph. The graph in Figure \ref{fig:opm}(a) is mapped to the ROPM in Figure \ref{fig:opm}(b), where ROSCs represent the graph vertices and the coupling blocks represent the graph edges. The illustration of the building blocks is simplified over detailed designs in Figure~\ref{fig:blocks} in order to highlight the mapping. Figure \ref{fig:opm}(c) illustrates ROPM converging to a solution through the minimization of the system energy after contention, where the 3-SHIL guarantees that solution at each ROSC is one of the $120 \degree$ spaced stable phases.  

\subsection{Probabilistic Computation of Potts Machines}
\label{sec:probabilistic_potts}

The proposed ROPM operates in a probabilistic manner, i.e. the initial state of oscillator phases determines to which ground energy state the system converges. These ground states can translate to the global minimum of the energy space which, in turn, would provide the exact (optimum) solution of the COP, or one of the local minima which would translate to a close-to-exact (quasi-optimum) solution. The ROPM can get stuck in local minima during the energy minimization. Consequently, running multiple iterations on the ROPM, starting from different initial states, and exploring more area in the solution space increases the chance of achieving better quality results. As an example, Figure \ref{fig:opm}~(d) shows of the solution space of the 8-node 3-coloring problem in Figure~\ref{fig:opm}(a).  The solution space entails $3^8$ possible phase combinations, with possible phases generically $[0 \degree, 120 \degree, 240 \degree]$ or any $120 \degree$ spaced phase values. Figure~\ref{fig:opm} marks sorted possible 3-phase spin combinations on the x-axis with their energy Hamiltonians shown in the y-axis. The highest energy states (i.e. the worst solutions) correspond to the 3 cases where all ROSCs are in phase with each other, therefore the graph is uni-colored. There are 12 global minima (optimal solutions) to this problem as marked by the blue dashed line in Figure~\ref{fig:opm}(d) corresponding to exact solutions with different 3-color permutations. The red dashed line marks one of the local optima (a quasi-optimum solution) corresponding to a ground state with some coloring mistakes lowering the accuracy.

\section{ Simulation Results}
\label{sec:simulations}

\begin{figure*}[t]
     \centering
     \begin{subfigure}[]{0.31\linewidth}
         \includegraphics[width=\linewidth]{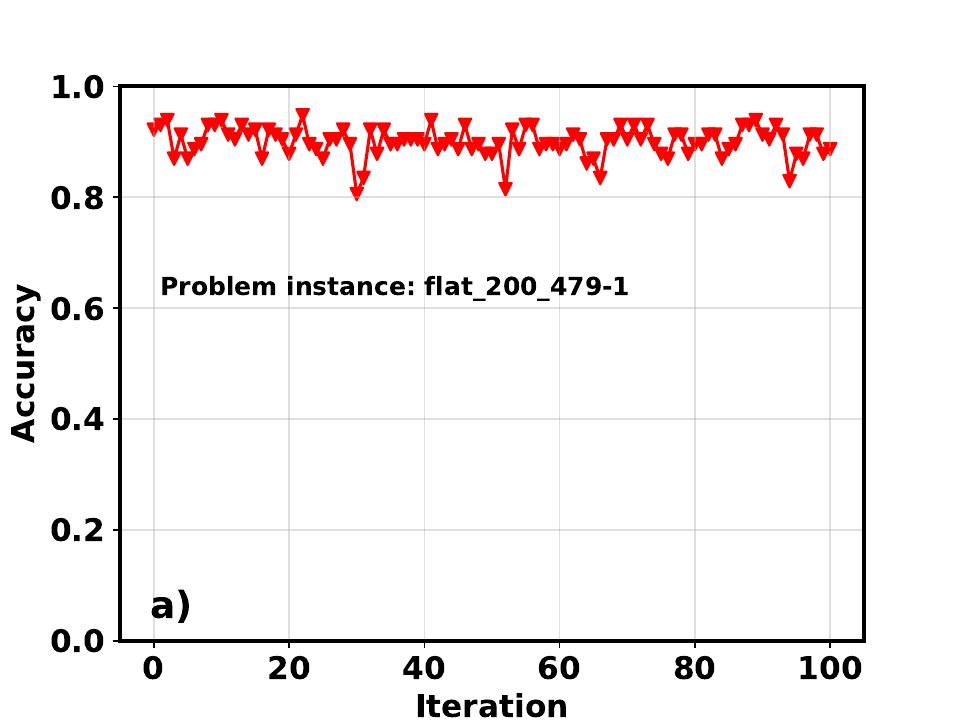}

    \label{fig:monte_carlo_single}
     \end{subfigure}
     \begin{subfigure}[]{0.31\linewidth}
         \includegraphics[width=\linewidth]{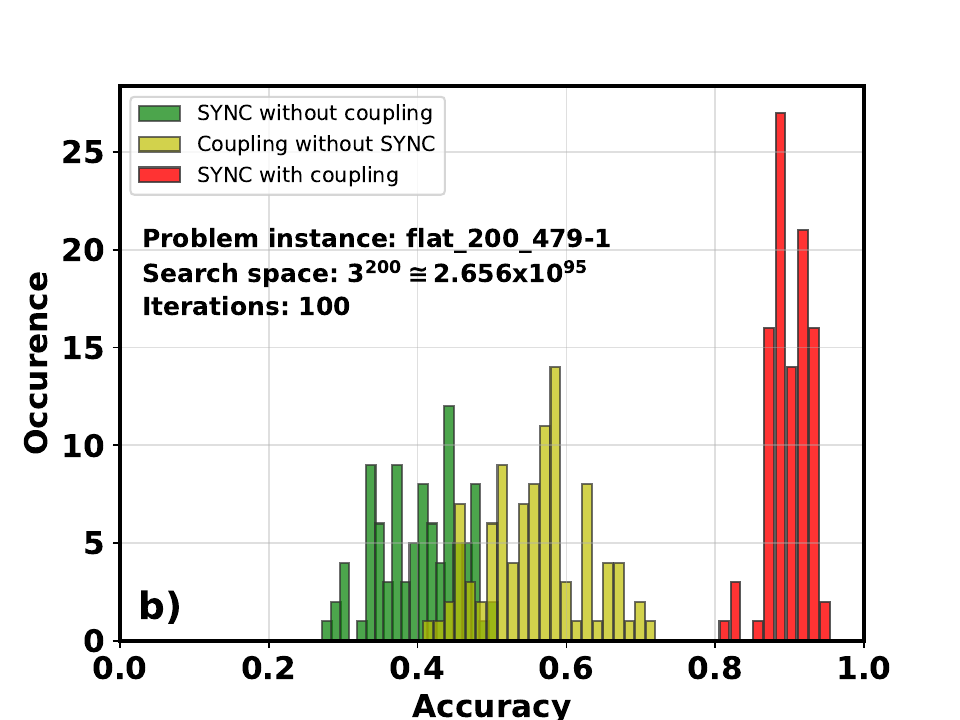}

         \label{fig:comp_monte_carlo}
     \end{subfigure}
     \begin{subfigure}[]{0.31\linewidth}
         \includegraphics[width=\linewidth]{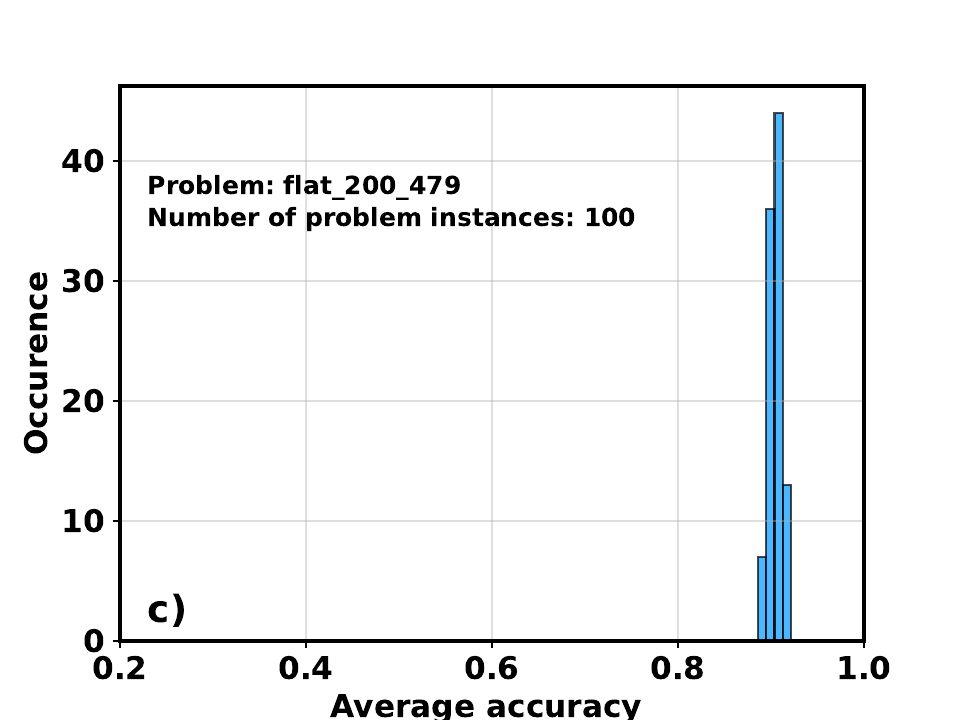}

         \label{fig:multiple_instances}
     \end{subfigure}
     \caption{Analysis of ROPM solution quality of results demonstrating a) accuracy levels of $\mathbf{flat\_200\_479}$-1 problem ran 100 times with randomized initial conditions b) Distributions of accuracies of 100 runs of solving $\mathbf{flat\_200\_479}$-1 problem on ROPM with both SYNC and couplings on (red), with SYNC on and couplings off (green), and with both SYNC and couplings off (yellow) c)~Averaged accuracy results obtained from running 100 different instances of the $\mathbf{flat\_200\_479}$ problem 100 times}
    \label{fig:results}
\end{figure*}

ROPM implementations of various sizes and parameters are simulated with Cadence Spectre using a 65nm technology node at 1~V operation. Flat 3-coloring problems from DIMACS SATBLIB \cite{satlib} graph coloring benchmark suite are mapped to the implemented ROPMs. The benchmark suite consists of 3-colorable graphs containing 100 different instances of 8 different problem sizes that are shown in Table 1. Two larger 3-colorable graphs are also randomly generated in order to observe the effectiveness of ROPM over graph sizes up to 10$\times$ larger than those in DIMACS SATLIB. The SYNC signal is generated as an ideal sine wave in the simulations. In practice, external voltage-controlled oscillators, or on-chip oscillators with good frequency scaling such as \cite{rotasyn} can be used.

\begin{table}[]
\centering
\caption{Performance metrics of ROPM on DIMACS SATLIB benchmarks}
\label{combstats}
\fontsize{8pt}{10.5pt}\selectfont
 \begin{threeparttable}
\begin{tabular}{||c |c |c | c| c|c|} 
\hline
 \textbf{Benchmark} & \textbf{Iter} &\textbf{Time*} & \textbf{Accuracy (Avg - Best)} &\textbf{Power}  \\ 
\hline
\hline
 flat\_30\_60-1  &100&  16  & 92\% - 96\% &17 mW  \\ 
\hline
 flat\_50\_115-1  &100 & 21 & 91\% - 95\% &32 mW  \\ 
 \hline
 flat\_75\_180-1  &100 & 25  & 90\% - 94\% &56 mW  \\ 
 \hline
 flat\_100\_239-1  &100 & 32 & 90\% - 95\% & 80 mW  \\
 \hline
 flat\_125\_301-1  & 100 &42  & 90\% - 94\% & 94 mW  \\
 \hline
 flat\_150\_360-1  &100 & 44  & 90\% - 93\% & 116 mW  \\
 \hline
 flat\_175\_417-1  &100& 49  & 90\% - 94\% &133 mW  \\
 \hline
 flat\_200\_479-1  & 100 &53  & 90\% - 94\% & 155 mW  \\
 \hline
\end{tabular}
 \begin{tablenotes}
            \small
             \item \raggedleft{* Number of ROSC cycles to solution}
        \end{tablenotes}
 \end{threeparttable}
\end{table}

\begin{table}[]
\centering
\caption{Performance metrics on random-generated graphs}
\label{randomstats}
\fontsize{8pt}{10.5pt}\selectfont
\begin{tabular}{|| c |c |c | c|c|} 
\hline
 \textbf{Graph} &\textbf{Iter} & \textbf{Time} & \textbf{Accuracy (Avg - Best) }  & \textbf{Power}  \\ 
\hline
\hline
 rnd\_1000  &100 & 60 & 90\% - 93\%  & 745 mW  \\ 
 \hline
 rnd\_2000  &100 & 74 & 89\% - 92\%  & 1.548 W  \\ 
 \hline
\end{tabular}
\end{table}

Coupled ROSC solvers in the literature are implemented in a specific topology, such as Hexagonal \cite{prob_fabric}, King's graph \cite{chris_kim_nat} or, all-to-all \cite{all-to-all}. In this work, due to most of the benchmark graphs being non-planar, no particular topology is used.
Instead, all problems are mapped to the corresponding ROPM implementations without any geometric constraints.  
ROPM circuits are reprogrammed and reinitialized 100 times (i.e. 100 iterations) for each benchmark problem (i.e. instance 1 of each problem size with trivial extension to other instances). Initial conditions are randomly changed each time to converge to different minimum energy states (different solutions). To randomize the initial states of the phases, ROSCs and couplings are turned on at randomized time instances.

As a metric for the quality of the results, previous work on Ising machines solving max-cut problems \cite{prob_fabric}, \cite{chris_kim_nat} use normalized cut-size of the solution. For the 3-coloring problem, the quality of results is expressed by counting the number of edges in the graph which satisfy the coloring rule. Normalized number of correctly colored neighbors indicates how close the converged solution is to an actual solution (global minimum). Exact solutions to the problems are computed using a generic software sat-solver serving as the baseline for accuracy calculations.

Table~\ref{combstats} demonstrates convergence speed (number of ROSC cycles to solution), power, and accuracy statistics for SATLIB 3-coloring benchmark problems of different sizes averaged over 100 iterations. Simulation results show that the increasing problem size has sub-linear effect on the convergence speed of the ROPM. There is a linear increase in power as the problem size increases. ROPMs implementing different size problems consume on average only 0.7 mW power per ROSC. 
Table~\ref{combstats} also demonstrates the superior performance of ROPM in accuracy: As the problem size scales, accuracy of the ROPM does not see any significant change, delivering solutions with 90\% accuracy on average for increasing problem sizes in the SATLIB benchmarks with the best accuracy at 93\% to 96\%.

The sustained accuracy, as well as solution time and linear power increase of ROPM on SATLIB benchmarks are very encouraging, motivating the investigation of larger graph coloring problems. To this end, 3-colorable graphs with vertex counts 1000, 2000 and edge counts 2682, 5662 are randomly generated. The same three performance metrics of proposed ROPM solutions of the three random graphs are shown in Table~\ref{randomstats}. Linear scaling of the power consumption persists as the number of nodes increases by 10$\times$ over SATLIB benchmarks. The accuracy achieved by the proposed ROPM is consistent at 89\% - 90\% on average, compared to 90\% - 92\% of SATLIB benchmarks. The sub-linear trend in the convergence speeds continues with the random graphs as well.

\subsection{Analysis of ROPM Performance Metrics}
\label{sec:analysis}

The ROPM solutions presented in Table~\ref{combstats} are analyzed further in order to investigate the factors that contribute to the reported performance metrics. The largest SATLIB benchmark problem, $flat\_200\_479$, is selected for the analysis.  The three goals of the analysis are 1)~To identify the impact of initial ROSC phases on the accuracy of the achieved solution, 2)~To identify the impact of SYNC (N-SHIL) and ROSC coupling on solution quality, and 3)~To identify the range of solution accuracy of ROPM on the 100 different instances of the problem  $flat\_200\_479$. 

The accuracy obtained from running 200-node 3-coloring benchmark problem  $flat\_200\_479$-1 100 times with different combinations of initial phases is shown in Figure~\ref{fig:results}(a). Each initial condition starts a potentially different path of energy minimization, converging the system to different local or global minima. 
Among 100 iterations, the highest achieved accuracy is 94\%, and the lowest is 80\%. In practice, after running the ROPM multiple times, the solution with the 94\% accuracy is to be picked being the best of all.  
It is also interesting to note that despite the 90\% accuracy on average, a global minimum (one of the possible exact solutions or accuracy of 100\%) is not reached in the 100 runs with randomized initial conditions. Mathematical simulation results in~\cite{seal_potts} also report that the introduction of N-SHIL only to an OPM does not guarantee that the exact solutions will be reached even for the smaller problems. Annealing schedules are reported to improve the results to achieve better, sometimes exact, solutions \cite{seal_potts}. 

\begin{table*}[t]
\centering
\caption{Comparison with prior work}
\label{table:comparison}
\fontsize{8.2pt}{10.2pt}\selectfont
 \begin{threeparttable}
\begin{tabular}{| c |c |c |c | c|c|c|} 
\hline
 & \textbf{This work} & \textbf{\cite{chris_kim_nat}}  & \textbf{\cite{rtwo_ising}} & \textbf{\cite{potts_nature}} & \textbf{\cite{optical_potts}} & \textbf{\cite{coherent_potts}} \\ 
\hline
 \textbf{Solver type} & Potts & Ising & Ising & Potts & Potts & Potts \\ 
\hline
 \textbf{Solved COP} & 3-coloring & Max-Cut & Max-Cut & 4-coloring & 3-coloring & 4-coloring \\ 
\hline
 \textbf{Technology} & CMOS 65nm GP & CMOS 65nm LP & CMOS 65nm GP&Optical \& Digital & Optical & Optical \\ 
 \hline
\textbf{Spins} & 2000 & 1998 & 2750  & 47  & 30  & 47  \\ 
 \hline
 \textbf{Average power} & 1.548 W & 42 mW & 17.48 W & DNR & DNR & DNR \\ 
 \hline
\textbf{Time to solution} &  11ns (74 cycles)  & 50ns (50 cycles)  & 10ns & 500 \micro s & DNR & DNR \\ 
 \hline
 \textbf{Accuracy*} & 83\%-92\% & 89\%-100\%  & 91\%-94\% & 50\% success rate** & 50\%-100\% & 20\%-100\% sucess rate** \\ 
 \hline
 \textbf{Baseline} & Exact solution & Tabu \cite{tabu} & SA  & Exact solution & Exact solution & Exact Solution\\ 
 \hline
\end{tabular}
 \begin{tablenotes}
            \small
             \item \raggedleft{* Accuracy is presented as the worst (if reported) and the best accuracy obtained w.r.t. the baseline over iterations}
            \item  \raggedleft{** Sucess rate refers to the number of times 100\% accuracy is achieved}
        \end{tablenotes}
 \end{threeparttable}
\end{table*}

Figure \ref{fig:results}(b) shows the distribution of the accuracy results of 100 runs of the ROPM mapping the $flat\_200\_479$-1 problem in 3 different settings. The histogram depicted by yellow bins shows the results achieved by rounding the ROSC phases to the nearest discrete phase through post-processing, with active couplings but with an inactive SYNC in the ROPM. Without 3-phase discretization provided by SYNC but with couplings only, the ROPM achieves accuracy levels under 60\% down to 20\% for the 100 iterations. Green bins show the results obtained with active SYNC in an uncoupled ROPM for each of the 100 iterations. SYNC in an uncoupled ROPM discretizes the oscillator phases randomly, achieving below 50\% accuracy. When the couplings are turned on together with the SYNC, as depicted by the red bins, accuracy levels are greatly increased to an average 90\% of the exact solutions, which highlights the importance of couplings and SYNC working in tandem to converge to better solutions. 

 Figure \ref{fig:results}(c) shows the distribution of the accuracy results of ROPM mapping 100 different instances of the $flat\_200\_479$ problem each averaged over 100 iterations. Each instance of the benchmark is the same size in terms of nodes and edges, but technically a different problem. The average accuracy of ROPM of solving the $1^{st}$ instance, with 100 different initial conditions, is 90\%, as reported in Table~\ref{combstats}.  The distribution in Figure~\ref{fig:results}(c) shows that the average accuracy results are never below 89\%, and reach higher accuracy levels of 91\% for some problems among the 100 instances of the benchmark circuit due to differences in problem instances (i.e. different connectivities being of different difficulties with different number of phase contentions). Observing these results on the largest benchmark problem gives confidence that the reported averages in Table~\ref{combstats} for 100 iterations of instance 1 of each benchmark problem size, are representative of all problem instances.

\subsection{Comparison of Results with Prior Work}
\label{sec:comparison}
A comparison with other Potts machines in the literature, implemented with different technologies, and an Ising machine implemented with the same (CMOS) technology is presented in Table~\ref{table:comparison}. 

Studies \cite{coherent_potts} and \cite{potts_nature} solve a 47 node 4-coloring problem with coherent Potts machines (CPM), the latter with an additional digital component and multi-stage operation. \cite{coherent_potts}  is reported to achieve the exact solution with 20\% to 100\% success rate depending on changing design parameters, whereas \cite{potts_nature} reports to achieve exact solutions with a 50\% success rate. Although the proposed ROPM is not able to reach exact solutions with the similar sized problem $flat\_50\_115$, the proposed ROPM, demonstrates results with a much larger number of spins up to 2000 with no degradation in accuracy, showing scalability. 
Reference~\cite{optical_potts} is another coherent Potts machine implementation using optical parametric oscillators solving an 30-node 3-coloring problem. The work in ~\cite{optical_potts} reports accuracy levels between 50\% and 100\% with an average of around 75\%, lower than the average 92\% accuracy achieved by the similarly sized problem ($flat\_30\_60$) with ROPM. Reference \cite{potts_nature} reports $500 \micro s$ run-time per stage, significantly higher than the 11 ns run time of the ROPM. References ~\cite{optical_potts} and \cite{coherent_potts} do not report the time to solution and neither of these studies report power consumption. In addition to the discussed metrics,
as discussed in Section~\ref{sec:pottsimplementation}, 
ROPM built with CMOS technology is more miniaturizable and low-cost compared to the CPMs.

The comparisons to the ROIM in~\cite{chris_kim_nat} and rotary travelling wave oscillator based Ising machine (RTWOIM) in \cite{rtwo_ising} are performed cognizant of the differences of the mapped problems, as well as the differences and the similarities of the architectures.  When the solution spaces of max-cut and 3-coloring for the same number of nodes are compared, 3-coloring has a much larger number of possible energy states i.e. $3^N$ to $2^N$, representative of the increased difficulty brought by increased number of spins. It is also important to note that the accuracy in~\cite{chris_kim_nat} is not with respect to the optimal solution but with respect to the solutions of a tabu method search solver~\cite{tabu}, which can be sub-optimal. The reported 83\% to 92\% accuracy (worst and best) of the proposed ROPM is with respect to the optimal solution of the 3-coloring problem, solvable by most SAT solvers, as performed in this work. Although the power budget of the ROPM is kept under 2~W, there is a relative increase in the power consumption compared to \cite{chris_kim_nat} which is partially caused by the increased ROSC frequency and the stronger couplings. The power budget of ROPM is 1.548~W simulated, significantly lower than that in \cite{rtwo_ising}, primarily thanks to utilizing ROSCs as opposed to RTWOs.      


\begin{figure}[]
 \centering
 \includegraphics[width=\linewidth]{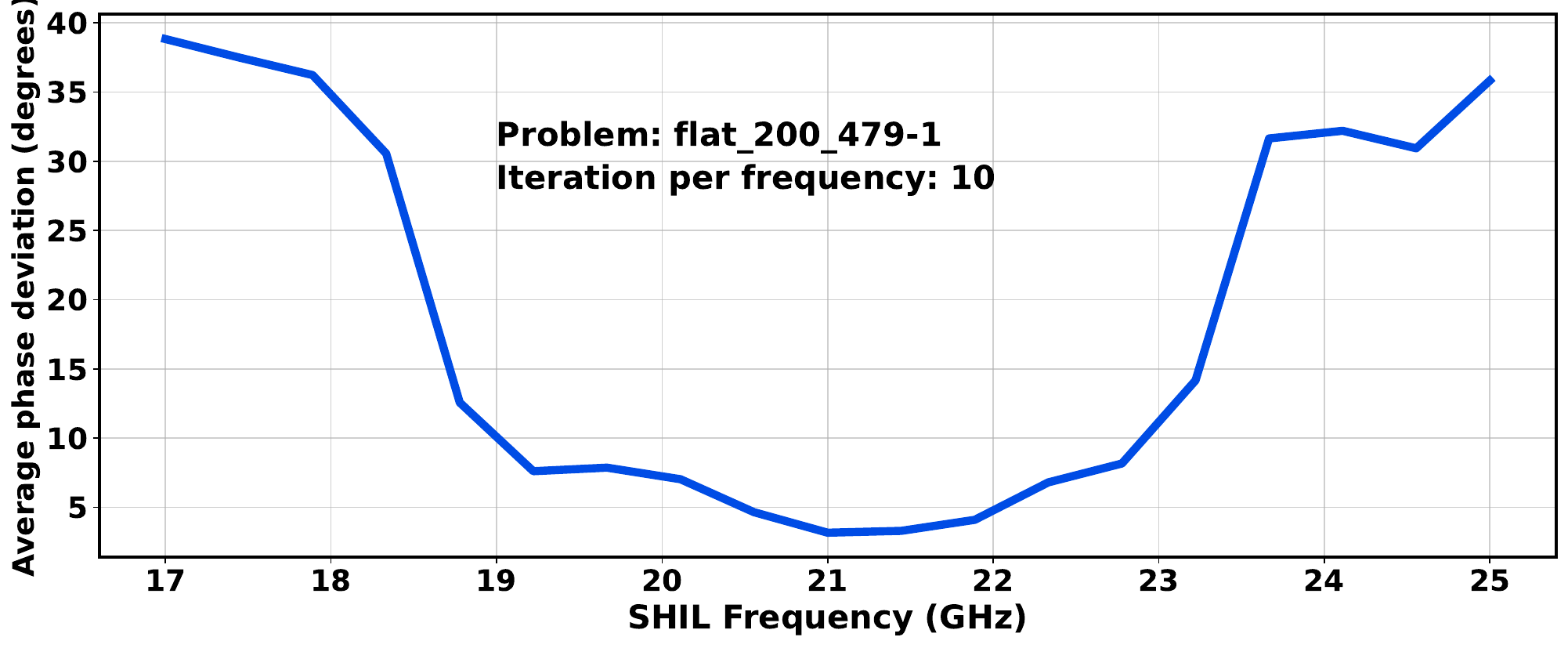}
    \caption{Robustness of three-phase discretization to shifts in SHIL frequency target 21 GHz, measured by the average deviation of each phase from the target phases}
\label{fig:freq_var}
\end{figure}
     
\subsection{Investigation on 3-SHIL Frequency}

For the 3-phase discretization of the  ROSCs at 7 GHz, $3^{rd}$ harmonic SHIL at a frequency of around 21 GHz (3 times the frequency of ROSCs) is required.  While the SHILs are modeled ideally in Spectre simulations, the silicon implementation will potentially lead to shifts in the target frequency of N-SHILs. To this end, a set of simulations are performed to investigate the impact of such a drift of the N-SHIL frequency from the target harmonic on the solution quality. Figure~\ref{fig:freq_var} shows the average deviation of all phases in the $flat\_200\_479$-1 problem averaged over 10 iterations for different SHIL frequencies. The exact $3^{rd}$ harmonic of 21 GHz discretizes the system with least amount of deviation. Frequencies within around 2 GHz range (10\%) of the exact $3^{rd}$ harmonic consistently achieve below $10\degree$ phase deviation. Considering the difficulty of generating high frequency of oscillation, the demonstrated tolerance to slight variations in the SHIL frequency makes the proposed ROPM highly practical in face of process and environmental variations.

\section {Conclusions}
\label{sec:conclusions}
This work presents a ring oscillator based Potts machine implementation obtaining multivalued spins from a single oscillator. Proposed ROPM is able to solve 3-coloring problems with 89\% and over accuracy, averaged over 100 iterations each, for various benchmark problems. The proposed ROPM can be extended to solve COPs with increasing spin-values leveraging the advantages of silicon with power, cost and size advantages against its counterparts built with different technologies.

\bibliographystyle{unsrt}
\bibliography{ref2}

\end{document}